\documentclass{aastex631}

\begin{document}

\title{SynCOM: An Empirical Model for High-Resolution Simulations of Transient Solar Wind Flows}

\author[0000-0002-5447-9964]{Valmir P. Moraes Filho}
\correspondingauthor{Valmir Moraes Filho}
\email{moraesfilho@cua.edu}
\affiliation{Department of Physics, Catholic University of America \\
620 Michigan Ave NE \\
Washington, DC 20064, USA}
\affiliation{NASA Goddard Space Flight Center \\
8800 Greenbelt Rd \\
Greenbelt, MD 20771, USA}

\author{Vadim M. Uritsky}
\affiliation{Department of Physics, Catholic University of America \\
620 Michigan Ave NE \\
Washington, DC 20064, USA}
\affiliation{NASA Goddard Space Flight Center \\
8800 Greenbelt Rd \\
Greenbelt, MD 20771, USA}


\author{Barbara J. Thompson}
\affiliation{NASA Goddard Space Flight Center \\
8800 Greenbelt Rd \\
Greenbelt, MD 20771, USA}

\author{Sarah E. Gibson}
\affiliation{University Corporation for Atmospheric Research}

\author{Craig E. DeForest}
\affiliation{Southwest Research Institute}



\begin{abstract}

The Synthetic Corona Outflow Model (SynCOM), an empirical model, simulates the solar corona's dynamics to match high-resolution observations, providing a useful resource for testing velocity measurement algorithms. SynCOM generates synthetic images depicting radial variability in polarized brightness and includes stochastic elements for plasma outflows and instrumental noise. It employs a predefined flow velocity probability distribution and an adjustable signal-to-noise ratio to evaluate different data analysis methods for coronal flows. By adjusting parameters to match specific coronal and instrumental conditions, SynCOM offers a platform to assess these methods for determining coronal velocity and acceleration. Validating these measurements would help to understand solar wind origins and support missions such as the Polarimeter to Unify the Corona and Heliosphere (PUNCH). In this study, we demonstrate how SynCOM can be employed to assess the precision and performance of two different flow tracking methods. By providing a ground-truth based on observational data, we highlight the importance of SynCOM in confirming observational standards for detecting coronal flows.

\end{abstract}



\section{Introduction}
     \label{S-Introduction} 

The Sun is enveloped by a layer of hot, magnetized plasma known as the solar corona. Due to the combination of a radial pressure gradient and magnetic forces, some plasma is able to overcome the Sun's gravitational attraction and expand outward \citep{Cranmer_2017}. This outward-flowing plasma, termed the solar wind, is a stream of magnetized plasma that originates from the Sun, interacts with the Earth's atmosphere, and forms the Heliosphere \citep{Sakao_2007,Dialynas_2017}. Early models of solar wind acceleration predicted steady-state spherically symmetric plasma outflow \citep{parker_1965}. However, high-resolution, high-cadence coronal observations have revealed the solar wind as a complex system with transient flow events such as jets, jetlets, type 2 spicules and other flow transients \citep{Viall_2010,Rouillard_2011,Cranmer_2017,Richardson_2018}. These transients are likely initiated by magnetic reconnection events in the lower corona \citep{Wyper_2022,Pellegrin-Frachon_2023,raouafi_2023}. Understanding these small-scale flows and their contribution to the large-scale solar wind is a central objective of modern solar and heliophysics \citep{Beedle_2022}.

New space missions are being developed to bridge the gap in our knowledge of these small-scale flows, such as the PUNCH mission (Polarimeter to Unify the Corona and Heliosphere), which aims to deliver high-sensitivity polarized and unpolarized brightness measurements from corona through the inner heliosphere \citep{DeForest_2022}. One of its primary scientific goals is to comprehend cross-scale physical processes, ranging from mesoscale turbulence to the development of global scale structures, thus providing a cohesive view of the solar corona and the wider solar system environment. Efforts directed towards this mission are concentrated on understanding the solar wind's structure. One of the scientific objectives of this mission is to generate a quantitative map of solar wind speeds throughout the Heliosphere, covering a range from 5 to 80 solar radii ($R_\odot$). The PUNCH mission will use algorithms to track the movement of transient features by calculating their velocities across consecutive images.

The task of tracking flows for the upcoming PUNCH mission depends on a set of algorithms designed to identifying the velocity of transient features in consecutive images. 
Predominantly, these tracking methods measure optical flow in sequential images, aiming to quantify the motion velocities in an image's brightness patterns due to the relative movement between the observer and the scene \citep{Horn_1981}. 
A commonly used method is Local Correlation Tracking (LCT), which uses two-dimensional local cross-correlation to identify the similarity between two regions in consecutive images. The motion is inferred from the displacement that maximizes the correlation between these images \citep{november_1988}. November and Simon used this method to identify the dynamics of solar granulation. Moreover, the detection of motion can be accomplished through the momentum tracking method, also known as 'balltracking' \citep{potts_2004,Attie_2009}, which is specifically designed to follow granules in photospheric images. This technique deploys 'balls' on the image surface that are drawn to local intensity minima, subsequently moving with the granules to observe their collective movement. A variation of this method, which uses magnetogram data and is referred to as 'magnetic ball tracking' \citep{attie_2015}, is currently being adapted for use in coronagraph images for the PUNCH mission.

Another method of velocity tracking in nearly parallel flows uses distance-time (DT) plots of brightness or running-difference brightness in the corona. Coronal images with radius on one axis and time on the other reveal moving objects as diagonal lines, whose strength indicates feature brightness and whose slope indicates feature speed.  These “J-maps” have long been used to quantify feature motion in the corona using direct visual analysis \citep{Sheeley_1999} or via the Hough transform \citep{Ballester_1994}, and applications of the method are reviewed by \cite{Llebaria_1999}. Variations of the DT image method include the surfing transform \citep{Uritsky_2009}. This integration method successfully smooths out random fluctuations, thereby retaining the initial temporal resolution of the image sequence and safeguarding the frequency distribution's consistency of the moving disturbances. This approach is thoroughly detailed by Uristky \textit{et al.}, (\citeyear{Uritsky_2013} and \citeyear{Uritsky_2023}). 


Combining DT image plotting with the Fourier transform can yield 'speed spectra' for regions in the corona, measuring the quantity of material traveling at specific speeds, regardless of its position \citep{deforest_2014}. This combination of the Fourier transform with DT image analysis techniques brings them into the realm of correlation methods, setting them apart from purely feature-based method. Through the incorporation of speed spectrum analysis, it becomes evident that the difference between DT and image-sequence approaches is distinct from that between correlation-based and feature-based techniques, with each offering unique perspectives on the dynamics of solar wind flows.

Each of these algorithms employ a different technique, such as tracking optical features, using correlation peaks or DT plots, but they all encounter similar issues, including background noise, stellar contributions, and transient feature motion. While these algorithms are able to generate quantitative data on solar wind flow speeds, they are not able to verify if these results are valid. They are also unable to assess their precision, highlighting the need for a model that provides a ground-truth reference. Such a model would allow for the verification and fine-tuning of these algorithms, enabling researchers to quantitatively measure the flow and determine which methods are most accurate and appropriate for specific conditions.

Advanced 3D MHD models, validated by observations, provide a comprehensive understanding of coronal structure, heating, and solar wind acceleration. Models such as WSA, AWSoM, among others, have significantly advanced the state-of-the-art in solar and heliophysics modeling. For instance, the Wang-Sheeley-Arge (WSA) model has enhanced predictive capabilities by incorporating real-time synoptic magnetic maps \citep{Wang_1990, Arge_2000}. The Alfvén Wave Solar atmosphere Model (AWSoM), part of the Space Weather Modeling Framework (SWMF), provides a comprehensive physics-based framework for space weather simulations \citep{Toth_2005, Toth_2012}, offering a global description of coronal heating and solar wind acceleration \citep{Sokolov_2013, van_der_Holst_2014}. The magnetohydrodynamic algorithm outside a sphere (MAS) model employs 3D MHD simulations in the inner heliosphere to ascertain the coronal magnetic field structure and the distribution of solar wind velocity, plasma density, and temperature, advancing the MHD equations over time with appropriate boundary conditions \citep{Linker_1999, Miki_1999, Riley_2011}. The GAMERA model has introduced high-resolution MHD simulations for complex geometries, improving the modeling of plasma environments such as the solar corona and planetary magnetospheres \citep{Zhang_2019}. Recent research has utilized 3D MHD simulations to explore potential mechanisms of magnetic switchbacks in the lower solar atmosphere \citep{Magyar_2021}. Even with these progressions, understanding the intricate behavior of solar wind is still difficult. Many physics-based models, while accurate and thorough, are typically resource-intensive and computationally heavy, which makes them impractical for real-time use or extensive simulations.

Although current physics-based models are praised for their high-resolution and temporal accuracy, they come with substantial computational costs. These models frequently rely on boundary conditions that do not precisely align with the initial goals of the observational missions or the specific characteristics they aim to observe. In the context of validating feature tracking codes, these models present considerable challenges. They are not ideal for this purpose as they fail to accurately represent the variability and complexities inherent in solar wind phenomena. Considering these limitations, there is a necessity for a more straightforward and efficient framework that could better facilitate the validation of feature tracking algorithms. This model would be advantageous for the community by providing a reliable ground-truth reference, allowing for more precise and effective assessment of these methods in various solar conditions.

The SynCOM (Synthetic Corona Outflow Model) is a data-driven model that offers an effective solution by focusing on the statistical properties of solar wind dynamics. This paper introduces the initial effort to tackle the challenges encountered by the solar observational community with a quick, high-resolution model directly constrained by solar observations. SynCOM generates high-resolution, statistically representative simulations rapidly, eliminating the need for detailed plasma equations. The primary benefits of SynCOM include its significantly faster performance and high spatial and temporal resolution, all accomplished on a single laptop under 15 hours. Additionally, SynCOM's outputs are much closer to real solar wind flows because they are derived from observed data. This makes it particularly valuable for interpreting observational data and validating flow tracking methodologies. By providing a specific target velocity for the assessment of these methods, SynCOM offers a practical and reliable ground-truth reference, benefiting the solar observational community by enabling more precise and practical assessments under various solar conditions.

This paper details the main design of SynCOM. In Section \ref{S-SynCOM}, we clarify our core motivations and outline the engine of SynCOM. In Section \ref{S-Test}, we discuss and evaluate a basic application of SynCOM, along with demonstrating the effectiveness of two distinct tracking methods. In Section \ref{S-Constraints}, we apply constraints to the model using measured solar data and assess the performance of the tracking techniques in retrieving the input ground-truth. In Section \ref{S-Community}, we describe a community initiative we have started to promote SynCOM. Finally, in Section \ref{S-Results}, we offer our concluding remarks and outline the initial steps towards our goal of creating a polarized data set using SynCOM and the FORWARD framework.

\section{SynCOM: The algorithm}
\label{S-SynCOM}


A key objective of SynCOM is to provide a synthetic dataset that can be used to train and test solar wind tracking algorithms. Current algorithms developed on observational data have the limitation that there is no "ground truth" to assess their accuracy. We have created a synthetic model with similar statistical spatial and temporal properties as observations and exact, known specification of local flow speed.  Analysis algorithm developers can then assess the performance of their methods and use the ground truth measurements from SynCOM to determine where and how the algorithms could be improved. 

The  IDL source code of the SynCOM software suite is available at https://zenodo.org/records/12674862 \citep{moraes_filho_2024}. The provided code enables high-resolution simulations of transient solar wind flows, including those described in this paper.

\subsection{The Propagating Gaussian blobs} 
    \label{S-Description}

The design of SynCOM was inspired by simulating the solar wind outflow emanating from the Sun. This led to the representation of solar wind clumps as radially expanding two-dimensional Gaussian perturbations. In this study, these structures are based on observations of plasma blobs that move outward from the corona, as documented by \cite{Sheeley_2009}. According to their findings, these irregularities begin around 3-4 $R_\odot$ from the center of the Sun as compact blobs of material, approximately 1 $R_\odot$ in length and 0.1 $R_\odot$ in width, separating from the tips of coronal streamers. The propagating Gaussian blob is a mathematical model of a transient propagating coronal feature represented in plane polar coordinates:
\begin{equation}
\label{eq:gaussian_blob}
    G(\theta, r) = G_0 \exp{\left(-\left[{\frac{(\theta-\theta_0)^2}{2L_\theta^2}}+{\frac{(r - r_0(t))^2}{2L_r^2}}\right]\right)} ,    
\end{equation}
where $\theta$ and $r$ are respectively the position angle ($\theta$, measured counterclockwise from the North pole at the solar limb) and the radial coordinates of the Gaussian blob, $\theta_0$ is the central $\theta$ position of the blob,  $r_0(t)$ is its central radial position evolving as a function of time $t$, $L_\theta$ and $L_r$ are the characteristic sizes of the blob in the $\theta$ and $r$ direction, correspondingly, and $G_0$ is the peak intensity of the blob. For the purpose of this paper, $\theta_0$ is chosen to be  time-independent, which results in the strictly radial propagation of the blobs.

As defined in equation (\ref{eq:gaussian_blob}), each Gaussian blob is a two-dimensional intensity array that changes over time and is determined by a four-element vector of adjustable parameters $[\theta_0, r_0, L_\theta, L_r]$ which specify the blob's position and shape.

\begin{figure}
    \centering
    \includegraphics[width=\linewidth]{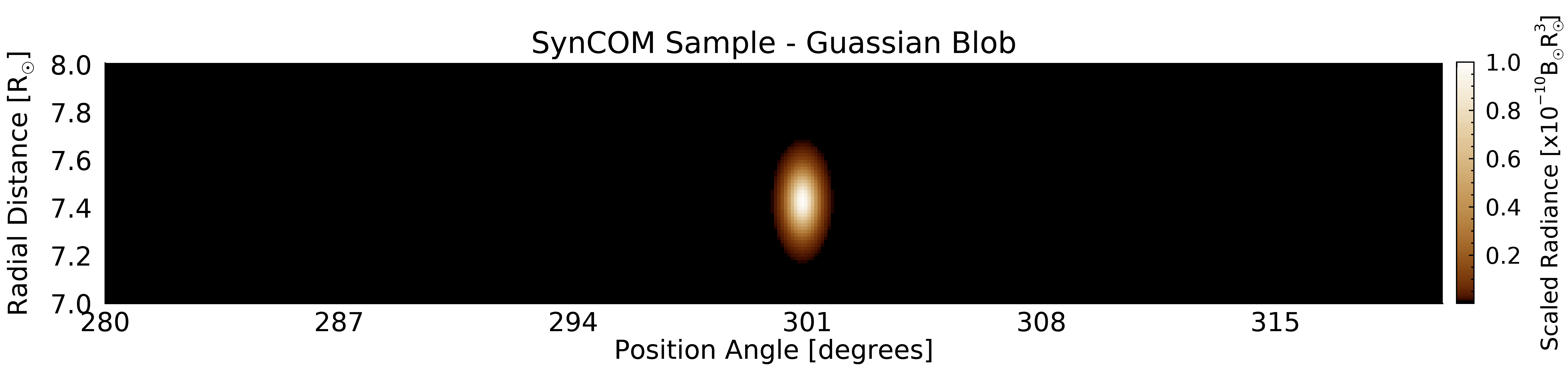}
    \caption{SynCOM blob example. This representation displays a Guassian blob sample, designed to emphasize its shape and size. In this instance, the brighter blob stands out in the image, appearing to be centered at angle 301 degrees and extending from 7.2 to 7.6 solar radii.}
    \label{fig:simple_sample}
\end{figure}

Figure \ref{fig:simple_sample} shows a Gaussian blob, originating from $\theta=301^\circ$ with its core at $7.4 R_\odot$. This paper uses the Heliocentric Radial Coordinates System as described by \cite{Thompson_2006}. The subsequent discussion will elucidate the methodology used by the model to generate a single simulated image.

We have developed a method to produce synthetic high-resolution images of solar wind density by simulating the motion of multiple Gaussian blobs within the solar corona. 
SynCOM includes a module that contains the essential parameters required to start its run, such as the number of blobs ($n_{blobs}$) being released, the number of pixels in each dimension of the image ($n_\theta, n_r$), and the initial launch position of each blob ($r_0$). Additional optional parameters that can provide other aspects of physical properties are also included in this structure, such as time cadence, pixel size, acceleration, peak brightness intensity, and noise level. These values were either obtained or estimated from a data set of the COR2 \citep{Howard_2008} instrument carried by the Solar Terrestrial Relations Observatory (STEREO) spacecraft \citep{Kaiser_2008}, see Section \ref{S-Constraints} for more details. 

The model includes an additional module that contains a database with profiles for the period, radius, and velocity for each $\theta_0$. These profiles can be derived from observed data or imported from external sources. 
This demonstrates the versatility of SynCOM. It can control how often features will appear, their sizes, and their position at a time t. For example, currently, the model uses a predefined function to determine the central location of the blobs for their initial launch,
\begin{equation}
\label{eq:blob_final_position}
    r_0(t)=r_B + V(\theta) t
\end{equation}
where $r_B$ is the blob's position at the observation boundary, $5 R_\odot$, and $V(\theta)$ is the radial velocity for a specific $\theta_0$. However, the future plan for SynCOM is to incorporate radial acceleration to the model, which will require a time-domain integration to equation (\ref{eq:blob_final_position}).

Each blob is then assigned with a $\theta_0$, that obtains its own set of variables from the database. For example, a blob will have its own $\theta_0$ with distinct values for velocity, period, radius, and initial radial and $\theta$ positions. 
Before creating a blob, $G(\theta_0,r_0(t))$, it is necessary to convert the velocity, radial position, and radius into pixel units. The model processes each blob at a time by generating a Gaussian blob with its unique set of variables. The blobs are periodically created and shifted radially to simulate their movement through the solar corona, creating multiple instances of each blob at regular intervals. These perturbations ensure that the blobs accurately reflect the quasi-periodic nature of the solar wind. The perturbed blobs are added to an initially empty image array, which accumulates the density distributions of all blobs over time. This process continues for all specified blobs, resulting in a comprehensive image that captures the observed visual dynamics of the solar wind, with comparable complexity to existing image data and a particular known underlying model. The image generated is saved as a FITS file, encapsulating the simulated solar wind densities at a given time $t$. Upon completion, the resulting collection of synthetic images, created over the designated time span, enables a detailed analysis and visualization of the transient dynamics of the solar wind.

\section{SynCOM: Performance tests}

In order to demonstrate and test SynCOM, we first use a simple "toy model" flow field (Section \ref{S-Test}) and then use a fit to solar data (Section \ref{S-Constraints}).

\subsection{Simple "Toy" Model} 
  \label{S-Test}

\begin{figure}
    \centering
    \includegraphics[width=\linewidth]{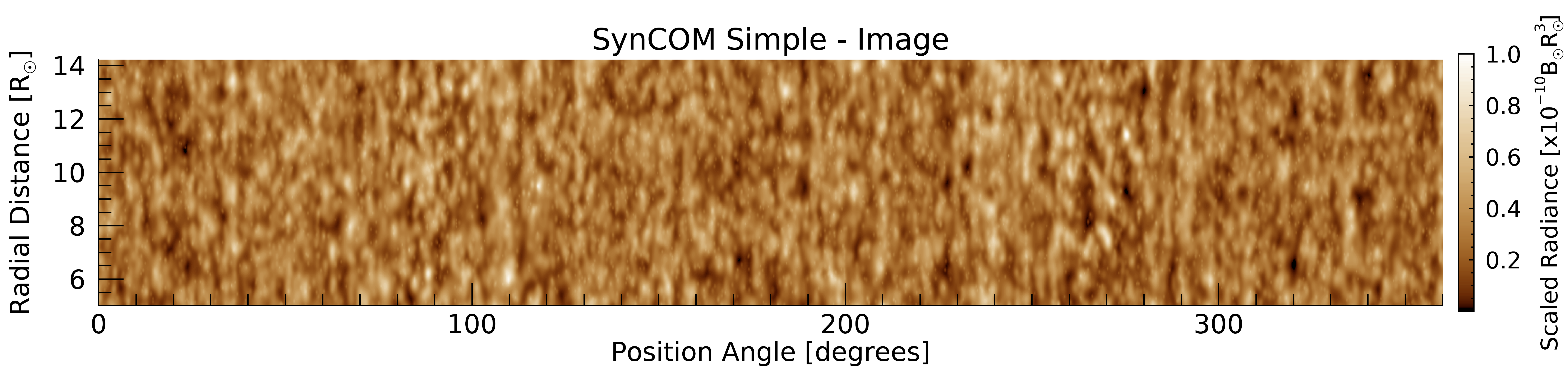}
    \caption{SynCOM simple numerical application featuring 5,000 blobs. This representation includes basic profiles for velocity, period, and blob radius inputs. The parameters utilized in these images are not physically significant but serve to evaluate the model's functionality.}
    \label{fig:simple_simulation}
\end{figure}

As a mathematical model, SynCOM is able to simulate coronal outflows using any type of velocity profile. This feature is also valid for any of its other input parameters, period, and blob radius, as shown in Table \ref{tab:parameters_simple}. To highlight this capability, we applied a simple distribution profile to each of our parameters within a toy model. This allowed us to generate SynCOM images with a defined flow-like dynamics that resembles coronal images; see Figure \ref{fig:simple_simulation}. For each of our parameters, we have utilized a different distribution profile. The velocity was attributed to a simple sinusoidal function that has its minimum and maximum as 150 and 350 km/s. The period and blob radius received a distribution where its minimum and maximum are, respectively, 1.5 and 4.5 hours, and 0.1 and 1.0 degrees. 

\begin{table}[ht]
\centering
    \begin{tabular}{ c|c|c|c|c }
    \hline
    Parameter          & Notation           & Min  & Max   & Units     \\
    \hline
    Number of Blobs    & $n_{blobs}$        & 0    & 5,000 & blobs     \\
    Time step          & $\Delta t$         & 0    & 70    & hours     \\
    Position angle     & $\theta$         & 0    & 360   & degrees   \\
    Radial distance    & $r$              & 5    & 14    & $R_\odot$ \\
    Period             & $T$                & 1.5  & 4.5   & hours     \\
    Radial velocity    & $V(\theta$)        & 150  & 350   & km/s      \\
    Blob radius        & $L_\theta,L_r$     & 0.1  & 1.0   & degrees     
    \end{tabular}
    \caption{Parameter boundaries for a SynCOM numerical test (see Figure \ref{fig:simple_simulation}). This test evaluates model capabilities with simple profiles.}
    \label{tab:parameters_simple}
\end{table}

\subsubsection{Velocity Profile}
\label{S-simple-velocity}

\begin{figure}
    \hspace{0.0 cm} (a)\\
    \includegraphics[width=\linewidth]{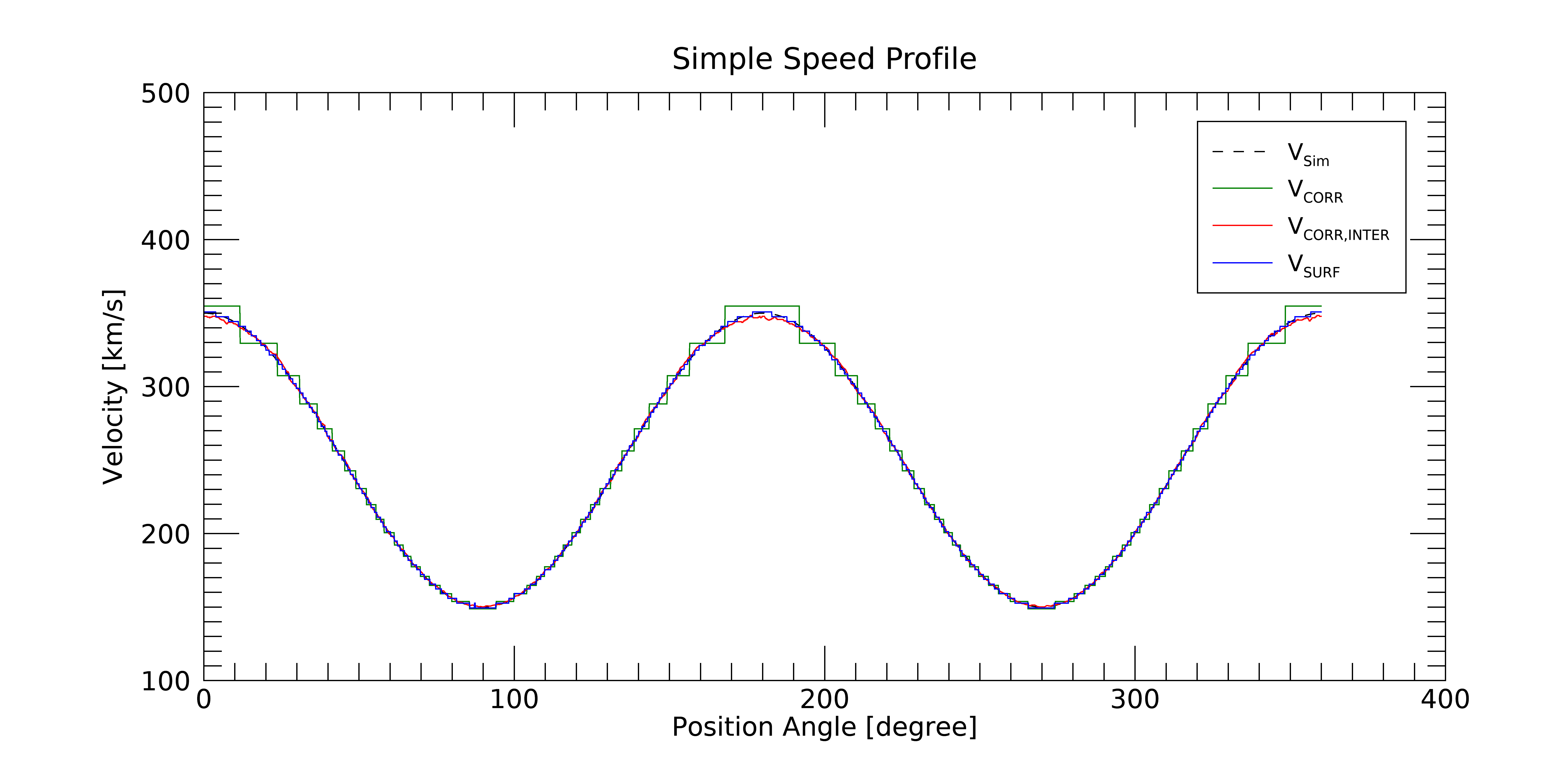}
    \hspace{0.0 cm} (b)\\
    \includegraphics[width=\linewidth]{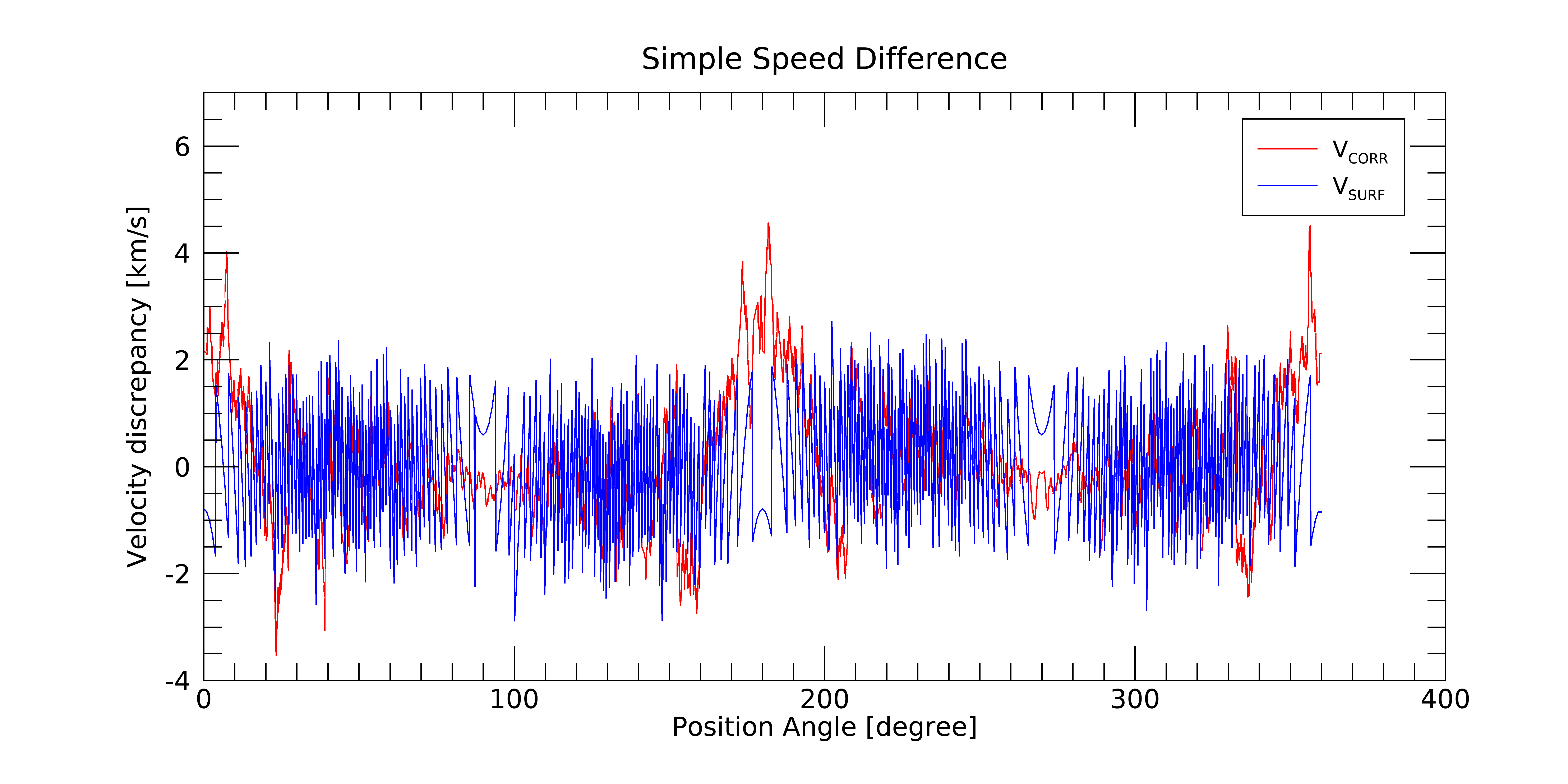}
    \caption{Analysis of a sinusoidal velocity profiles. \textbf{(a)} Displayed in \textit{black}, the Simulated velocity profile, characterized by a sinusoidal wave, was used to produce SynCOM images; shown in \textit{blue}, the Surfing velocity profile, which uses the surfing transform to integrate the DT plot along a dynamic spatiotemporal path encompassing various possible speeds \citep{Uritsky_2013}; in \textit{green}, the Correlation velocity profile, which determines a time delay to maximize the correlation peak; and in \textit{red}, the Interpolated Correlation velocity profile, an enhanced version of the correlation method where the peak is derived from the second derivative of a third-order polynomial. \textbf{(b)} The differences between simulated and measured speeds. Highlighted in \textit{red}, the speed difference for the correlation method; and in \textit{blue}, the speed difference for the surfing transform.}
    \label{fig:profiles}
\end{figure} 

The simple, toy model velocity profile we imposed was measured using two different tracking algorithms, the time-domain correlation (TDC) flow tracking and the surfing transform (ST) techniques. Each of these flow-tracking technique provides one velocity value for each positional angle, producing a velocity profile for the entire corona.

In the TDC, we define its motion as the displacement that maximizes the spatially localized cross-correlation between two time series separated by a sampling time delay $\Delta t_\theta$ that is shorter than the lifetime of the tracers in the scene. The spatially localized cross-correlation $C(\Delta t_\theta)$ is a function of the one-dimensional vector displacement $\Delta t_\theta$ between the time series:
\begin{equation}
    C(\Delta t_\theta)=\frac{\left< I(\theta_0,t,r_1) \cdot I(\theta_0,t +\Delta t,r_2)\right>_{t_{\theta}}}{\overline{\sigma}_1 \cdot \overline{\sigma}_2}
\end{equation}
where $C(\Delta t_\theta)$ is defined in terms of the time series $I(\theta_0,t,r_1)$ and $I(\theta_0,t+\Delta t,r_1)$ that sample the scene at the two intervals $t$ and $t+\Delta t$ for a given $\theta_0$, $\overline{\sigma}_1$ and $\overline{\sigma}_2$ are the variances for the respective time series correlations. For each pair of time series, we evaluated our velocity, in km/s, with a fixed spatial window of $1 R_\odot$:

\begin{equation}
    V_{CORR} \approx \frac{\Delta r_{fixed} }{\Delta t_{peak}}
\end{equation}

The second flow tracking methodology that we used for our tests, the ST technique, is designed to identify propagating disturbances in DT plots with low signal to noises ratios \citep{Uritsky_2013}. It has been used successfully for measuring parameters of remotely observed flows and waves in the solar corona \citep{Uritsky_2009, keiling_2012, Uritsky_2013, uritsky_2021, kumar_2022, raouafi_2023, mondal_2023}. The ST algorithm is based on an analysis of a velocity-dependent surfing signal $S(t, u)$ obtained by averaging the DT plot along an ensemble of spatiotemporal tracks characterized by different assumed propagation velocities $u$ and starting times $t$. The dynamic range of the ST signal is maximized when the assumed velocity matches the true propagation velocity such that the algorithms ``surfs'' the propagating wave front. This optimization criterion is used to calculate the ST-based flow speed:

\begin{equation}
    V_{SURF} = \arg \, \max_{t} \, S(t,u)
\end{equation}

See \cite{Uritsky_2013, uritsky_2021} for more details on the ST method.

\subsubsection{Recovering Velocity Profile from SynCOM images}
\label{S-Simple-Performance}

\begin{figure}[ht]
    \hspace{0.0 cm} (a)\\
    \includegraphics[width=\linewidth]{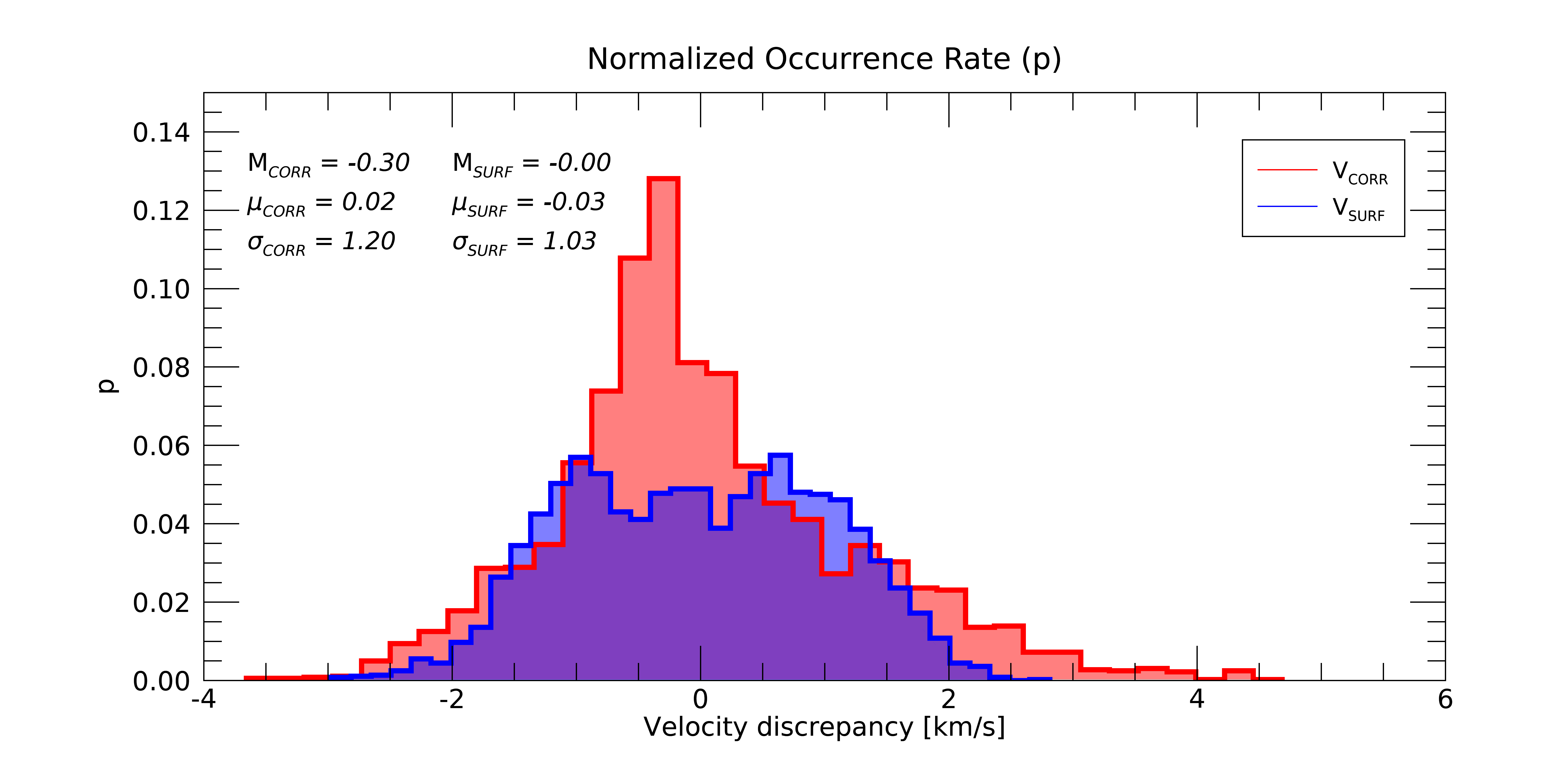}
    \hspace{0.0 cm} (b)\\
    \includegraphics[width=\linewidth]{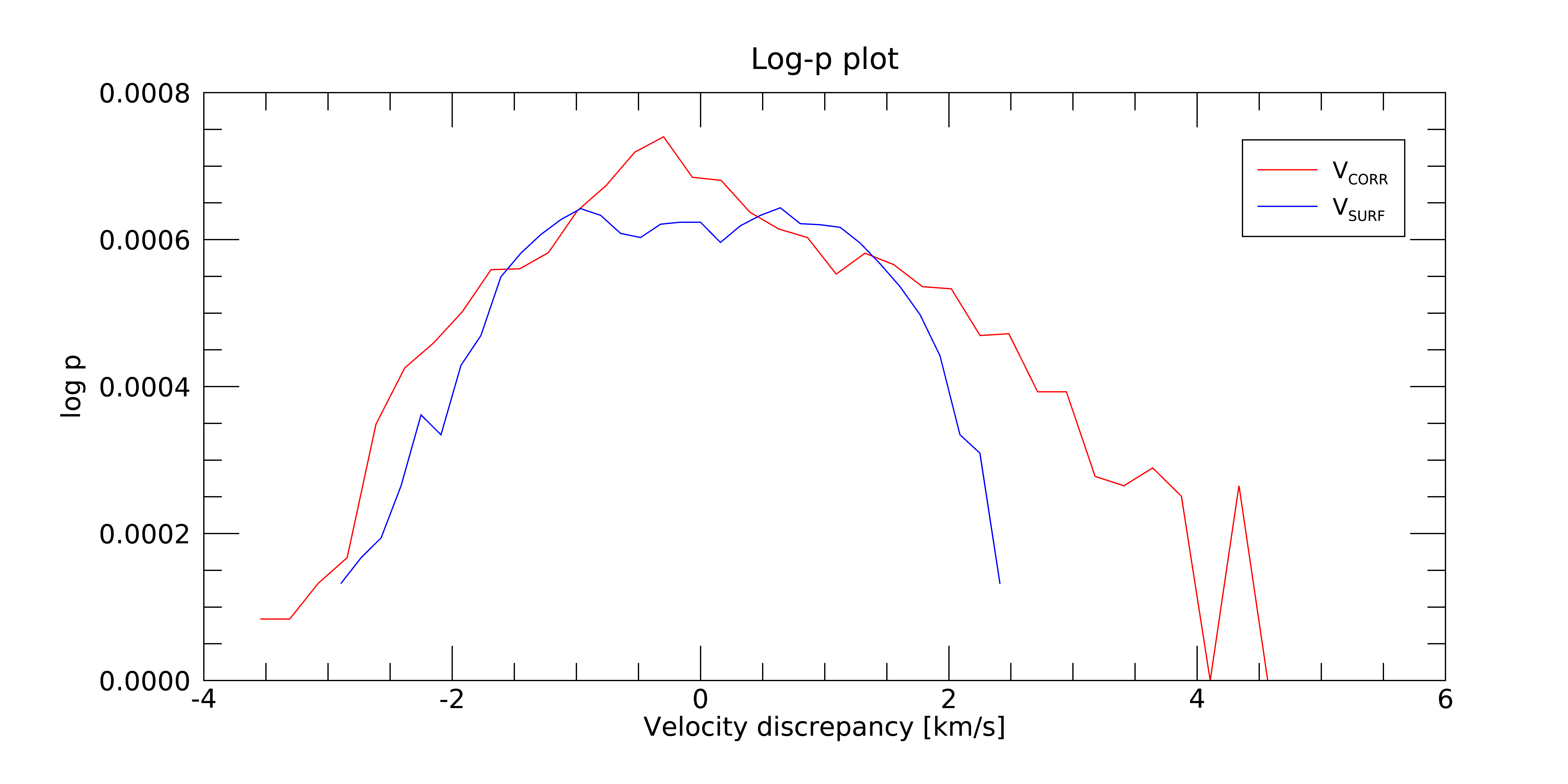}
    \caption{Analysis of velocity measurement errors. \textbf{(a)}, Probability distribution histograms for the speed difference ($\delta v$) using two approaches, Surfing transform in \textit{blue} and Correlation Method in \textit{red}. Each histogram includes calculations of the mean, standard deviation, and median. \textbf{(b)}, a semi-log graph showing the log of p vs. $\delta v$. Here, $\delta v$ denotes the disparity between the simulated and measured speeds.}
    \label{fig:histogram}
\end{figure}

The techniques chosen effectively described the velocity profiles in a simple SynCOM simulation, capturing a unique velocity value for each $\theta_0$. The ST demonstrated more accurate precision in the fitting of velocity profiles compared to the TDC, as demonstrated in Figure \ref{fig:profiles}(a). Although the correlation technique did not achieve the same accuracy as the ST, it yielded values that approximated the expected results. This difference revealed that the TDC did not accurately estimate the maximum peak, which required improvements. This improvement incorporated an interpolation term into the TDC, using the maximum peak from the initial correlation algorithm to develop a polynomial that accurately represents the neighboring data points. A third-order polynomial was utilized to determine the velocity values. The interpolated correlation analysis, improved by interpolation, showed better accuracy, as illustrated by the red line in Figure \ref{fig:profiles}(a). Henceforth, this interpolated technique will be referred to simply as the TDC. 

While both flow-tracking techniques achieved an acceptable velocity profile, Figure \ref{fig:profiles}(b) highlights the discrepancies in speed between the simulated and observed values for each method; the TDC now includes interpolation. This shows that the TDC exhibited greater variability compared to the ST, despite the latter experiencing more fluctuations. These deviations represent the measurement errors associated with each technique. The occurrence rate of these errors is visualized more effectively using a histogram, as shown in Figure \ref{fig:histogram}(a). The histogram for the TDC aligns closely with a normal distribution, as anticipated for random (nonsystematic) errors. It is relatively narrow, which corroborates the accuracy of the algorithm, suggesting its applicability to more intricate flow patterns, as seen in Figure \ref{fig:histogram}(b). In contrast, the histogram for the ST exhibits a narrow but non-Gaussian distorted configuration, attributed to the nonlinear characteristics of the ST \citep{Uritsky_2023} and the limited range of scanning velocities used in detecting the surfing resonance. 

The toy model demonstrates that SynCOM can accurately replicate flow-like dynamics using the simple sinusoidal profile. With regard to flow-tracking techniques, they managed to precisely capture the sinusoidal dynamics present in SynCOM. Although the ST offers greater accuracy, it requires significant computational time. In contrast, the TDC, particularly with the inclusion of interpolation, provides a faster yet comparably precise alternative. These findings highlight the adaptability of SynCOM in both flow-tracking techniques.


\subsection{Realistic Model based on Observational data} 
  \label{S-Constraints}

\begin{figure}
    \hspace{0.0 cm} (a)\\
    \includegraphics[width=\linewidth]{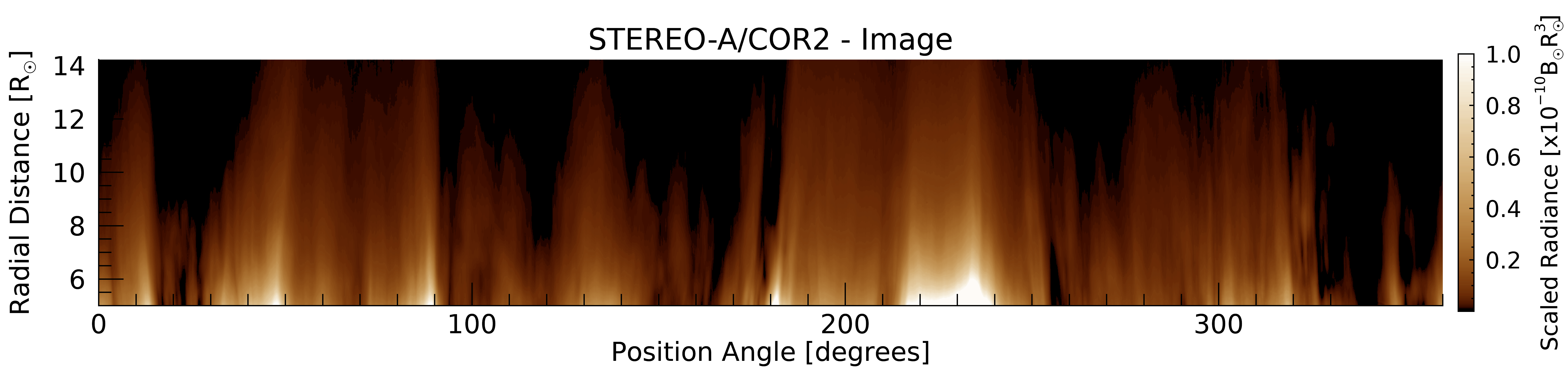}
    \hspace{0.0 cm} (b)\\
    \includegraphics[width=\linewidth]{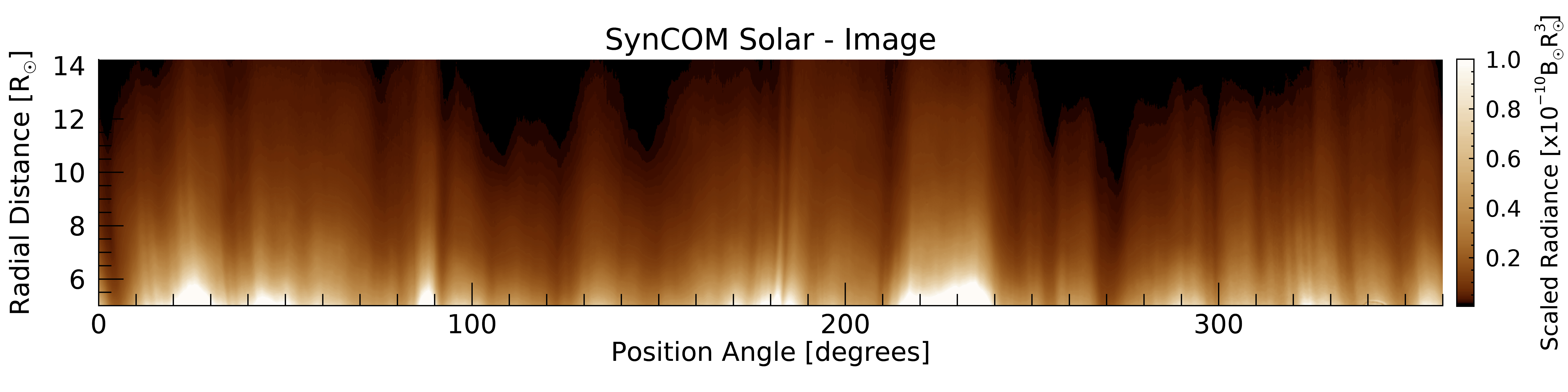}
    \hspace{0.0 cm} (c)\\
    \includegraphics[width=\linewidth]{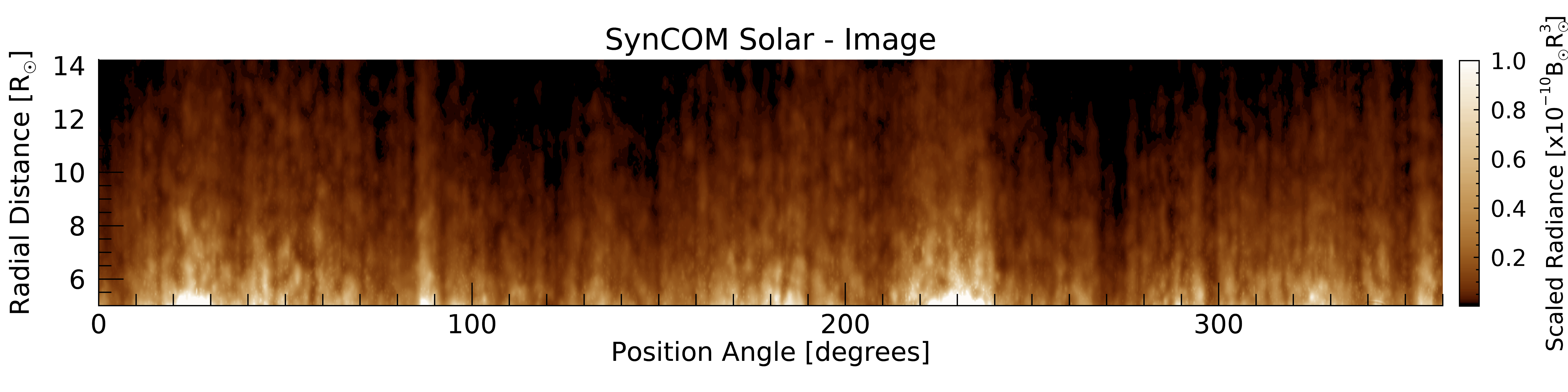}
    \caption{Comparison between a COR2 and a SynCOM simulated image; (a) Training data set from STEREO-A/COR2 instrument. (b) SynCOM simulation using solar constraints with 5,000 blobs. The blob sizes are doubled, making the image appear more natural compared to COR2 images. (c) SynCOM simulation using solar constraints with 5,000 blobs. The smaller blob sizes highlight its fine-scale structures.}
    \label{fig:syncom_solar}
\end{figure}

In the previous section, the efficiency of SynCOM was demonstrated using a basic model for velocity versus position angle, along with parameterized model inputs. SynCOM was shown to handle various distributions of velocity, period, and radius profiles, as presented in Table \ref{tab:solar_parameters}. This section will concentrate on creating a simulation incorporating solar parameters, as depicted in Figure \ref{fig:syncom_solar}(b). To achieve this, we will derive the velocity profile from the training data to create a synthetic flow-like structure that reflects the characteristics of the observed corona outflow dynamics. Subsequently, we will use flow-tracking algorithms to verify that the simulated velocity profile extracted from the COR2 dataset is accurately captured. Additionally, the luminosity profile from the COR2 images will be incorporated as an additional feature into SynCOM. Finally, we will evaluate the performance of SynCOM based on how effectively the flow-tracking methods can recover the ground-truth (simulated) velocity profile.

\begin{table}
\centering
    \begin{tabular}{ c|c|c|c|c }
    \hline
    Parameter          & Notation           & Min  & Max   & Units     \\
    \hline
    Number of Blobs    & $n_{blobs}$        & 0    & 5,000 & blobs     \\
    Time step          & $\Delta t$         & 0    & 70    & hours     \\
    Position angle     & $\theta$         & 0    & 360   & degrees   \\
    Radial distance    & $r$              & 5    & 14    & $R_\odot$ \\
    Period             & $T$                & 1.5  & 4.5   & hours     \\  
    Radial velocity    & $V(\theta)$        & 196  & 709   & km/s      \\
    Blob radius        & $L_\theta,L_r$     & 0.1  & 1.0   & degrees     
    \end{tabular}
    \caption{Parameter boundaries for SynCOM in a solar simulation, based on Figure \ref{fig:syncom_solar}(c). The constraints of the solar model use COR2 data for training. Currently, only the velocity profile is derived from the COR2 data. Future work will include period and radius profiles.}
    \label{tab:solar_parameters}
\end{table}
 
\subsubsection{STEREO-A/COR2 data set}
  \label{S-Training}

\begin{figure}
    \centering
    \includegraphics[width=\linewidth]{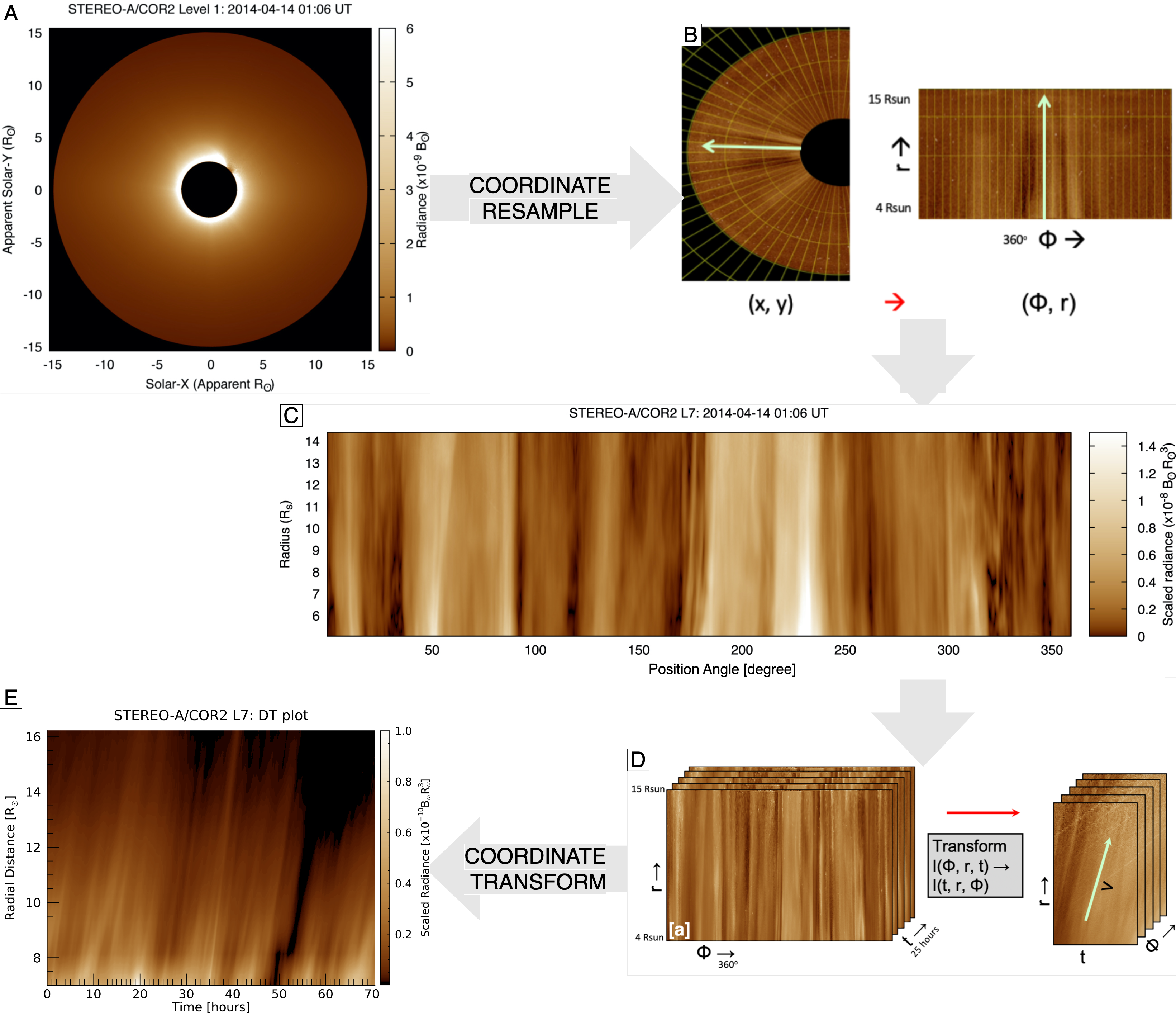}
    \caption{Summary of process for data preparation. \textbf{(A)} Level 1 (L1) generated using SECCHI software. This dataset includes a 2048 x 2048 pixel image, unpolarized and calibrated in units of average solar radiance along the x and y axes. Source: \cite{deforest_2018}. \textbf{(B)} Illustration of coordinate resampling. The system of image coordinates is converted from Cartesian to polar coordinates. \textbf{(C)} Level 7 (L7) image, representing one of the final phases of the data preparation sequence, following considerable modifications and filtering of the initial images. These images underwent several stages of processing, including adjustments for exposure-time fluctuations, transformations into polar coordinates, resampling, normalization, despiking to remove stars, and temporal smoothing. The image shows the solar corona outflow in polar coordinates $(\phi, r)$, with radiance adjusted by $r^3$. Source: \cite{deforest_2018}. \textbf{(D)} Representation of the coordinate transformation for J-maps. These maps were produced by selecting a thin sector or segment from a series of solar images and aligning these segments vertically in time order. J-maps serve to illustrate the kinematic progression of solar events across time and space. \textbf{(E)} STEREO-A/COR2 DT map. This map illustrates the temporal and spatial progression of features. The inclines on the map indicate the velocity of the features for that particular segment.}
    \label{fig:L1_resample_L7}
\end{figure}
  
To refine the model constraints, we employed observational data from the STEREO-A/COR2 instrument. During a special three-day campaign from 2014 April 14 00:00 UT through 2014 April 16 23:59 UT, the instrument collected deeper exposures of the corona. Under normal operational conditions, the instrument collects synoptic exposures that last 6 seconds at 20-minute intervals. However, for this specific campaign, exposures were extended to 36 seconds at 5-minute intervals, achieving an approximate 2.4-fold decrease in photon counting noise per image and a nearly 10x reduction over each 20-minute period \citep{deforest_2018}. Additionally, the total noise in each image also incorporated a minor component called 'compression noise', which arises from the image compression process prior to downlink transmission to Earth. The employment of prolonged exposures coupled with tailored accumulation strategies culminated in the generation of a distinctive deep-field data set.

Figure \ref{fig:L1_resample_L7} illustrates the SECCHI data preparation process. The Level 1 (L1) data set of 2048 x 2048 pixel unpolarized images, calibrated as mean solar radiance (Figure \ref{fig:L1_resample_L7}(a)). An ad hoc background subtraction produced the L2 data set, influenced by coronal structures like streamers. Saturation due to shutter fluctuations led to the removal of affected frames. The data were converted to polar coordinates and resampled to 3600 x 800 pixels for the L3 set (Figure \ref{fig:L1_resample_L7}(b)). Normalization produced the L4 set, from which extraneous stellar contributions were removed to form L5. To reduce motion blur, temporal averaging within a dynamic reference frame created L6. The brightness gradients were restored and a baseline value was subtracted, enhancing the radiance of the characteristics of the L7 set (Figure \ref{fig:L1_resample_L7}(c)). For comparison with L2, the data was reconverted to focal plane coordinates, forming L8 with improved contrast and feature visibility; see \cite{deforest_2018} for detailed methodology.
To facilitate our analysis, we create DT plots shown in Figure \ref{fig:L1_resample_L7}(d). These maps, used for monitoring the evolution of solar phenomena, are made by stacking vertical slices of solar images in sequence. The data for training SynCOM came from analyzing these plots, as depicted in Figure \ref{fig:L1_resample_L7}(e).

According to the research by \cite{deforest_2018}, the L7 and L8 images revealed a considerable amount of fine-scale structure at all position angles and radial distances. These observations include radial plasma optical flow and inhomogeneous density. Throughout the three-day period of this campaign, six coronal mass ejections (CMEs) were identified, moving more rapidly than the background flow, except for one slow CME. This slow CME, covering position angles from $100^\circ$ to $170^\circ$, took place from April 14 at 14:36 UT to April 16 at around 4:36 UT and was only noticeable due to the coronal depletion that followed. Additionally, familiar blobs were seen, such as those at position angles from $250^\circ$ to $270^\circ$ between April 14 at 00:00 UT and April 15 at 03:00 UT, following a CME eruption from the previous day.

\subsubsection{Intensity Profile}    
      \label{S-intensity} 


To enhance the fidelity of our simulation relative to the COR2 data, it is necessary to modify the visual appearance of the blobs in the simulation. Specifically, a feature must be introduced to modulate the luminosity as a function of distance. We have chosen to compute the mean and standard deviation across all COR2 data and then apply these statistical metrics to each pixel in a SynCOM image. This modification has produced a series of simulated images that closely emulate the intensity profile observed in the COR2 images. 

The collection of SynCOM images is denoted by $I_{t,i,j}$, where $t=1,2,...,T$ and each image $I_t$ is composed of a matrix with pixel values $I_{t,i,j}$, where i and j represent the pixel coordinates and T denotes the total number of images. The methodology to replicate the intensity profile is defined by: 
\begin{equation} 
    I'_{t,i,j} = \left( I_{t,i,j} * \sigma_J(i,j) \right) + \overline{J}(i,j)
\end{equation} 
where $I'_{t,i,j}$ represents a pixel in a SynCOM image that replicates the intensity profile of the COR2 images; $\overline{J}(i,j)$ and $\sigma_J(i,j)$ are the mean and standard deviation for each pixel position $(i,j)$ throughout all COR2 data images, with pixel values $I_{t,i,j}$. The mean and standard deviation are computed as follows:
\begin{equation} 
    \overline{J}(i,j) = \frac{1}{T}\sum^T_{t=1} J_{t,i,j} 
\end{equation} 
\begin{equation} 
    \sigma_J(i,j) = \sqrt{\frac{1}{T} \sum_{t=1}^{T} (J_{t, i, j} - \overline{J}(i, j))^2}
\end{equation},

This approach is a "reverse normalization" process. It reintegrates the statistical parameters observed in the COR2 data into our simulated images. This ensures that each pixel's intensity values are closely aligned with those observed in the COR2 data.

\subsubsection{Velocity Profile}
\label{S-velocity}

\begin{figure}
    \hspace{0.0 cm} (a)\\
    \includegraphics[width=\linewidth]{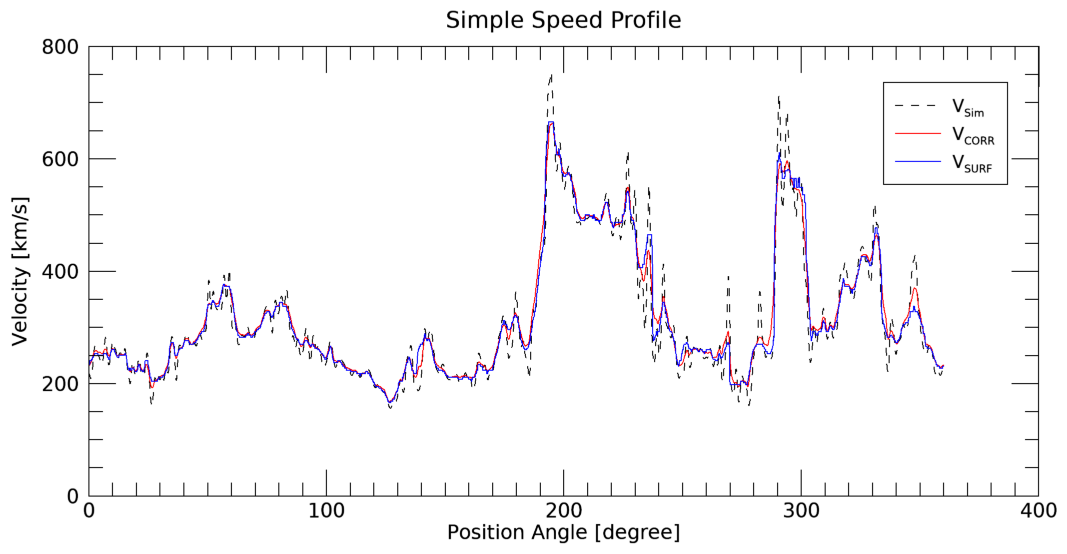}
    \hspace{0.0 cm} (b)\\
    \includegraphics[width=\linewidth]{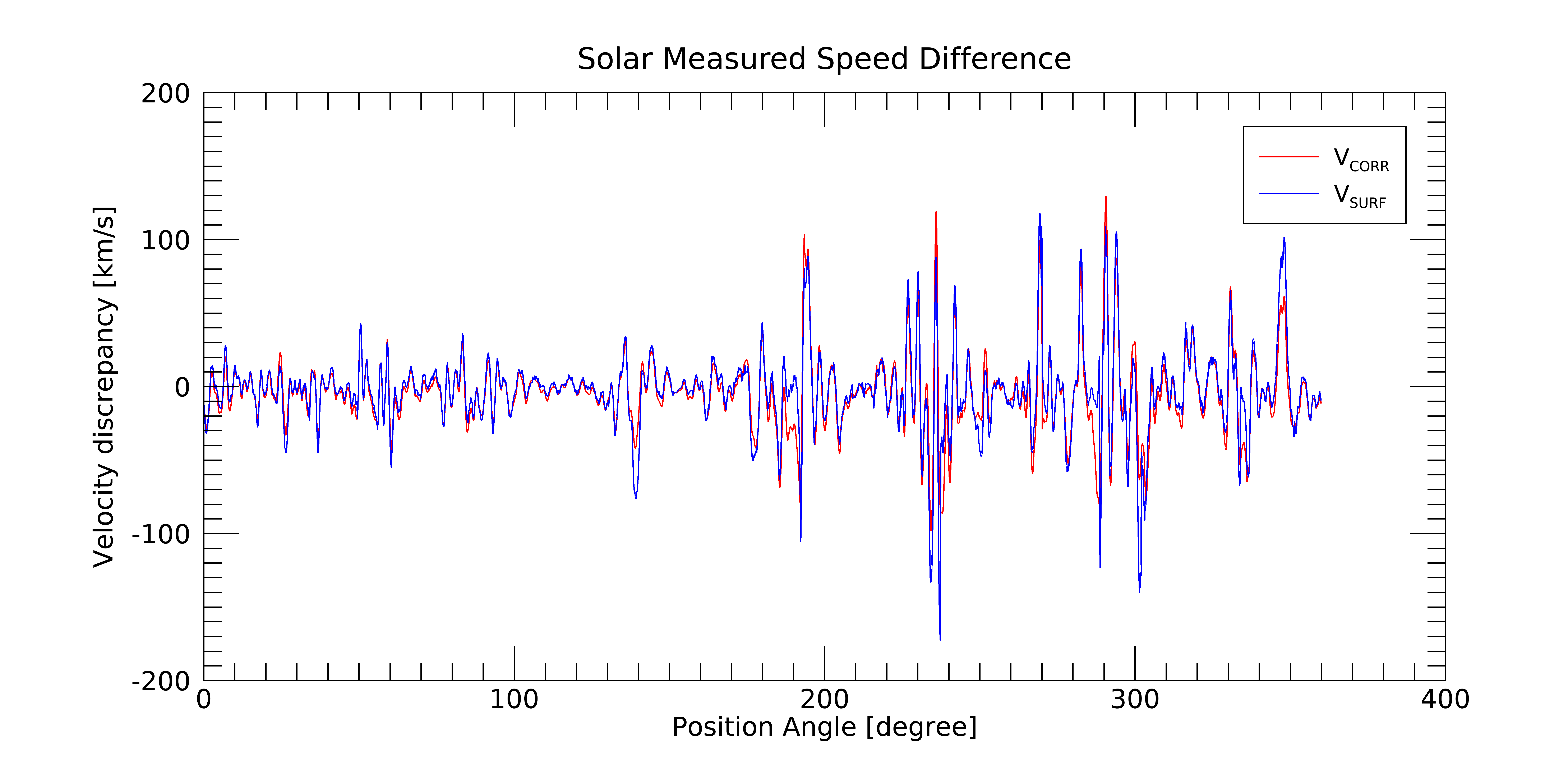}
    \caption{Simulated profile and its analysis. \textbf{(a)} Speed profile analysis of solar parameters. In \textit{black}, Simulated COR2 speed profile, a speed profile measured from COR2 data using correlation method and used to create SynCOM images with solar parameters; In \textit{blue}, Surfing transform speed profile; And in \textit{red}, Correlation speed profile. \textbf{(b)} Illustration of difference between simulated and measured speeds. In \textit{red} the speed difference for the correlation method; and in \textit{blue} the speed difference for the surfing transform. }
    \label{fig:solar_profile}
\end{figure}

SynCOM was designed to accept any type of velocity profile. This was demonstrated in Section \ref{S-simple-velocity}, where we used a simple sinusoidal speed profile, as shown in Figure \ref{fig:profiles}(a). This allowed us to create SynCOM images with predefined dynamics that reassembled solar images. The same methods described in Section \ref{S-simple-velocity} will be used to test the speed profile in the solar constrained SynCOM images that we created. 

We measured the speed profile used to render the images with solar parameters using the correlation method. The profile in Figure \ref{fig:solar_profile}(a) was measured from the COR2 data and smoothed to create our simulated images. Furthermore, both tracking methods were able to compute a speed profile for the simulated images.

\subsubsection{Recovering Velocity Profile from SynCOM images}
\label{S-Performance}
\begin{figure}[ht]
    \hspace{0.0 cm} (a)\\
    \includegraphics[width=\linewidth]{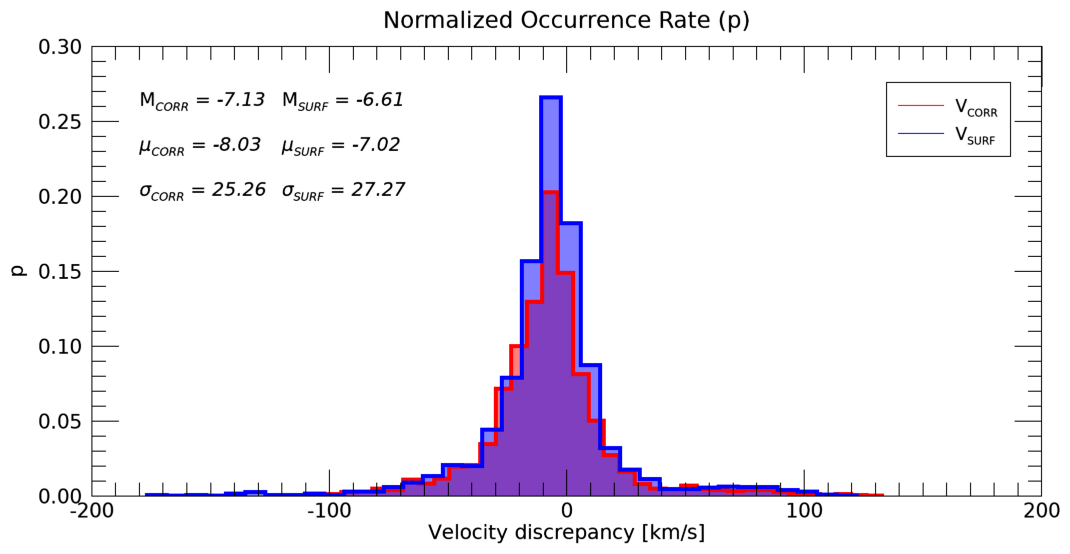}
    \hspace{0.0 cm} (b)\\
    \includegraphics[width=\linewidth]{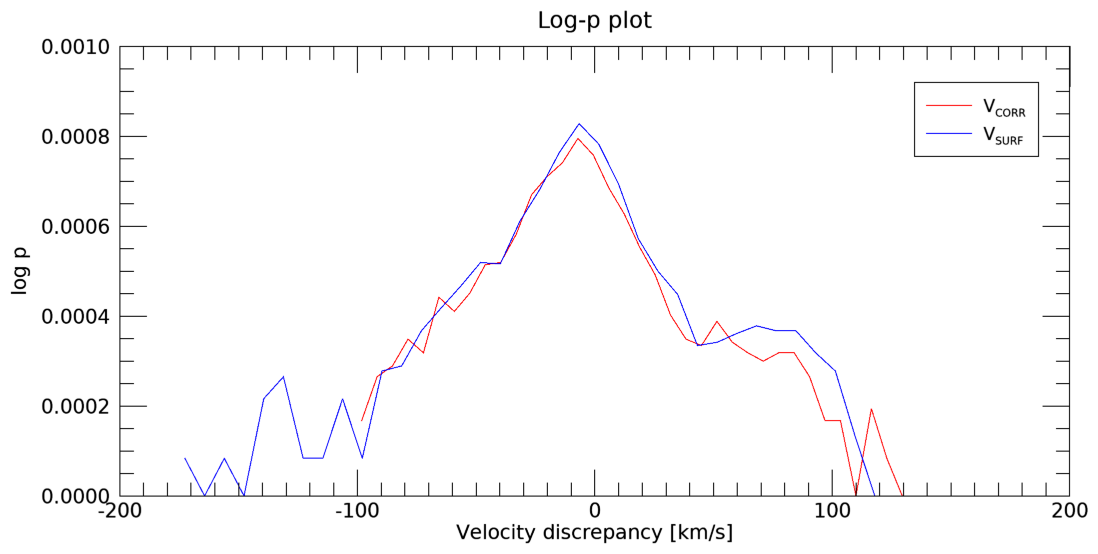}
    \caption{Analysis of velocity measurement errors. \textbf{(a)}, probability distribution histograms for the speed difference ($\delta v$) using two approaches, Surfing transform in \textit{blue} and Correlation Method in \textit{red}. Each histogram includes calculations of the mean, standard deviation, and median. \textbf{(b)}, a semi-log graph showing the log of p vs. $\delta v$. Here, $\delta v$ denotes the disparity between the simulated and measured speeds.}
    \label{fig:solar_histogram}
\end{figure}

As in Section \ref{S-simple-velocity}, we employed the same techniques to measure the velocity profiles, which closely matched the simulated velocity profiles used for the simulated image. Generally, each algorithm was able to closely estimate the solar constraint at various positions $\theta_0$, as depicted in Figure \ref{fig:solar_profile}(a). In addition, the TDC method provided speed measurements at various altitudes. In terms of measurement errors between the simulated and measured speeds, both methods showed similar fluctuations in numbers. Specifically, the ST exhibited fewer but sharper peaks, while the TDC method displayed a large pattern of mild variations, as shown in Figure \ref{fig:solar_profile}(b).

Histogram analysis confirms that the TDC method aligns with the normal distribution of stochastic errors, as seen in Figure \ref{fig:solar_histogram}(a). The narrow distribution emphasizes the precision and suitability of the method for complex flow dynamics. The ST histogram shows a non-Gaussian aberrant profile, consistent with simple simulation evaluations. The technique remains effective within set limits, with an error distribution described as 'well behaved' and reflective of simple test results, as shown in Figure \ref{fig:solar_histogram}(b).

The constrained model produced synthetic images that closely matched the observed COR2 data, shown in Figure \ref{fig:syncom_solar}. This validation underscores SynCOM's potential as a reliable tool for solar wind analysis. By accurately replicating solar wind phenomena, SynCOM enhances our understanding and ability to predict solar wind behavior.


\begin{figure}[ht]
    \centering
    \includegraphics[width=\linewidth]{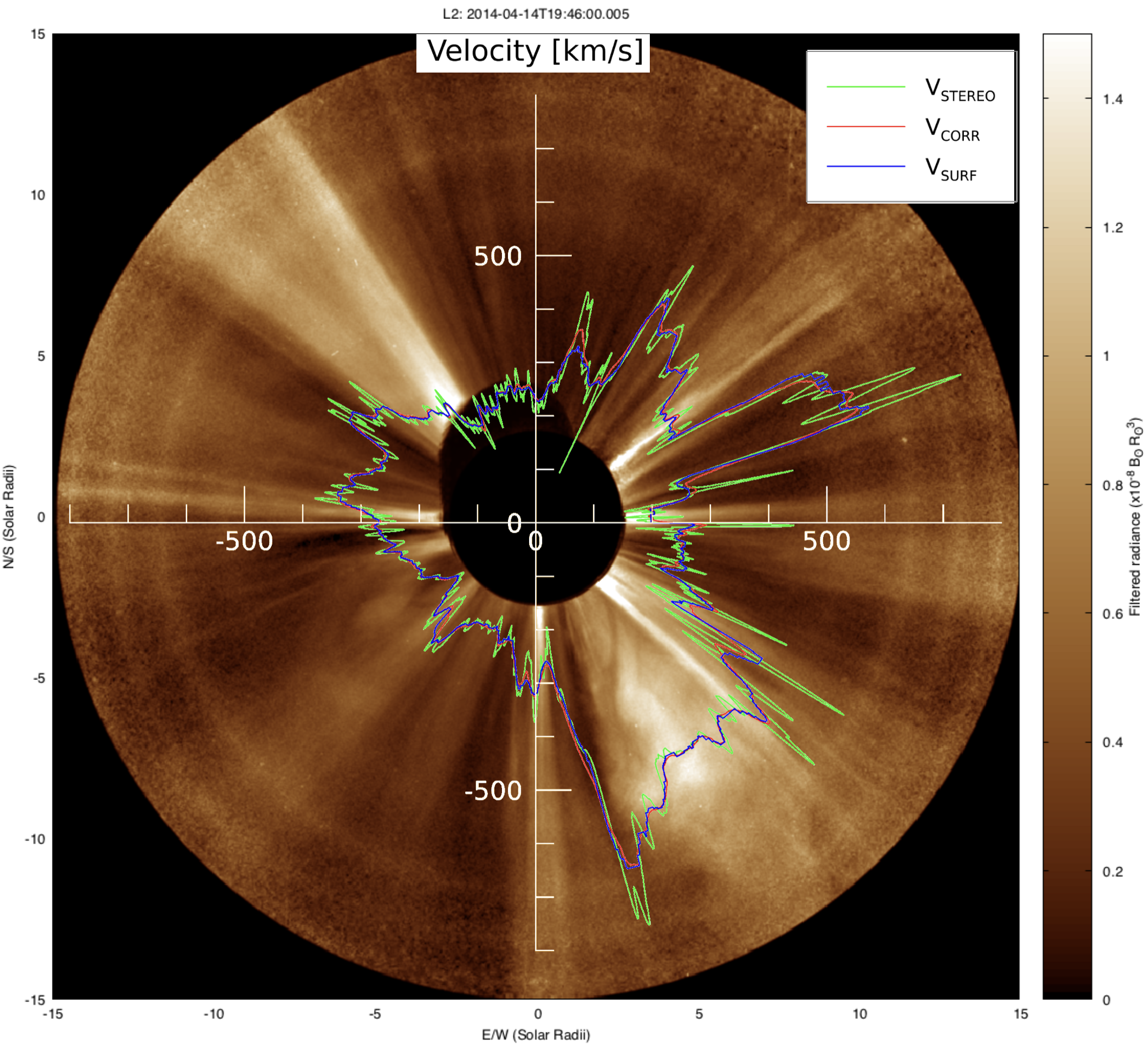}
    \caption{Measured Solar Velocity plot. This figure shows the velocity distribution around the Sun at different position angles. The observed velocity anomalies between position angles may indicate the presence of CMEs in the STEREO data, as mentioned in \cite{deforest_2018}, due to the setup of the velocity measurement method. The superimposition of the velocity plot on the coronagraph image emphasizes the slow CME at the velocity kink.}
    \label{fig:velocity_distribution_rect}
\end{figure}

\begin{figure}[ht]
    \centering
    \includegraphics[width=\linewidth]{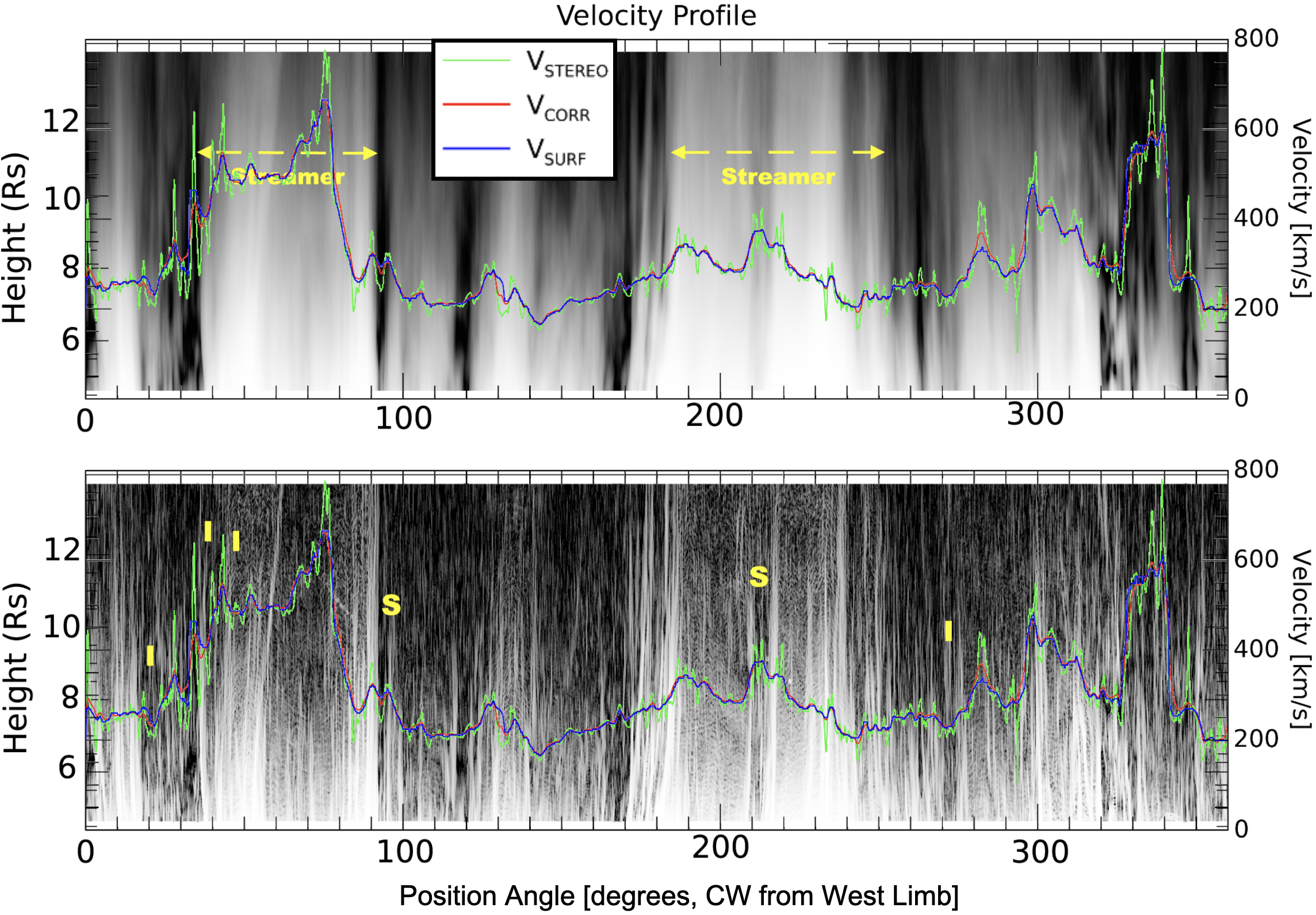}
    \caption{Fine-scale structures associated with the velocity distribution. \textbf{Top}: An L7 snapshot with several features of interest annotated. The velocity distribution overlay indicates that some anomalies in the velocity profile align with the highlighted features. \textbf{Bottom}: The same image processed using the Sobel operator (refer to \cite{deforest_2018} for more details) to emphasize brightness gradients (edges). The velocity plot uncovers the fine-scale structures in the highlighted image, matching the minor irregularities observed in the solar velocity profile. Background image adapted from from \cite{deforest_2018}. Note: this figure has the polar angle plotted as specified by \cite{deforest_2018}, which is measured clockwise from the West limb. In all other sections of this paper, we adopt the convention of measuring the polar angle counterclockwise from the North pole.}
    \label{fig:velocity_distribution_polar}
\end{figure}

The velocity profile derived from the simulated data, when transformed into polar coordinates, shows anomalies between position angles, which may indicate the presence of CMEs in the STEREO data, as noted by \cite{deforest_2018}. The alignment of the velocity profile with the STEREO image in rectangular coordinates, as shown in Figure \ref{fig:velocity_distribution_rect}, further supports this conjecture. These anomalies were propagated due to the configuration of the TDC method, which detects small variations in the quasi-periodic behavior observed in the solar wind. As seen in Figure \ref{fig:velocity_distribution_polar}, the small variations closely align with those in the velocity profile. It should be noted that this figure has the polar angle plotted as specified by \cite{deforest_2018}, which is measured clockwise from the West limb. In all other sections of this paper, we adopt the convention of measuring the polar angle counterclockwise from the North pole.

\section{SynCOM: Flow Tracking Challenge}
\label{S-Community}

The SynCOM Flow-Tracking Challenge was designed to involve the community in the testing of SynCOM and evaluating the performance of additional flow-tracking algorithms. It is available on the Catholic \cite{FlowTracking} website. First presented at the 2022 SHINE workshop, the challenge originally included just a mini-challenge involving DT plots with a predefined velocity. Currently, it encompasses a two-stage challenge: the initial version and a comprehensive main version that replicates real solar images with a specific velocity profile. The website provides files in FITS, SAV, and MOV formats, accommodating all flow trackers to evaluate their techniques and provide feedback to us, thereby improving the SynCOM's performance and also validating different algorithms.

\section{Conclusions and Future Work} 
\label{S-Results}


We presented the concept of SynCOM, explained its functionality, demonstrated some applications, and now invite contributions and new applications. SynCOM has achieved its primary objective of creating a high-resolution synthetic dataset to act as a "ground truth" for solar wind flow tracking techniques. Continuous enhancements are made to SynCOM, including its integration into the FORWARD framework \citep{Gibson_2016}. We hope that this integration will provide a fresh perspective on coronal outflows.

We developed SynCOM to provide a synthetic dataset for training and testing solar wind tracking algorithms. Current algorithms lack "ground truth" for accuracy assessment. A synthetic model with similar spatial and temporal properties offers exact measurements, allowing developers to evaluate and improve their algorithms. The framework is composed of a set of modules that provide the parameters for the run. The database it uses to determine its spatial and temporal scales can be derived from any source.

Furthermore, we have tested SynCOM using a simple sinusoidal function as its velocity profile. Using two different flow-tracking techniques on this simple application, we successfully measured the same sinusoidal function employed in the simulation. 
Our primary goal involved utilizing a dataset from the STEREO-A/COR2 instrument to refine SynCOM based on solar parameters. This implementation uses a velocity profile derived from the dataset using the TDC method. The flow-tracking algorithms successfully derived a velocity profile that closely matched the simulated one. 
Figure \ref{fig:velocity_distribution_rect} represents the velocity distribution throughout the corona, demonstrating that the direct measurements from SynCOM and STEREO are consistent. This graph highlights the occurrence of velocity kinks at particular $\theta_0$. This illustration can be associated with the presence of CMEs as noted in the research by \cite{deforest_2018}.

In conclusion, SynCOM provides a practical and efficient solution by focusing on the statistical properties of solar wind dynamics. This paper has demonstrated the initial steps in overcoming challenges faced by the solar observational community through a quick, high-resolution model constrained by solar observations. SynCOM's ability to generate high-resolution, statistically representative simulations rapidly—without relying on detailed plasma equations—brings significant advantages, including faster performance, high spatial and temporal resolution, and on-device execution. Its computational efficiency and simplicity surpass traditional physics-based models. The outputs of SynCOM closely match real solar wind flows as they are derived from observed data, making it invaluable for interpreting observational data and validating flow tracking methodologies. By providing a specific target velocity for assessment, SynCOM serves as a practical and reliable ground-truth reference, enabling more precise and practical evaluations under various solar conditions.

Next, we aim to integrate SynCOM with the FORWARD framework \citep{Gibson_2016}. This will incorporate physical parameters into our statistical model, enabling the production of polarized coronal images. Such images will not only enhance the predictive capabilities of SynCOM, but will also provide a synthesized data set that reflects the observations projected from the PUNCH mission. This collaboration is expected to yield new insights into the structure and behavior of the solar corona.




\begin{acknowledgments}
The authors thank the Southwest Institute for hosting the STEREO-A/COR2 data prepared by Craig DeForest.  We also thank Nicki Viall, Raphael Attie, Anna Malanushenko, and Elena Provornikova for fruitful discussions of the modeling techniques and the solar wind flows and structure. V.P.M.F. has been supported through the Partnership for Heliophysics and Space Environment Research (NASA grant No. 80NSSC21M0180). V.P.M.F. extends his gratitude to the NSF NCAR's Advanced Study Program's Graduate Visitor Program (GVP) for his fellowship in 2023.
\end{acknowledgments}

\bibliography{sample631}{}
\bibliographystyle{aasjournal}



\end{document}